
\magnification=\magstep1
\hsize=15.7truecm \vsize=23.4truecm
\baselineskip=6mm
\font\fontP=cmcsc10
\font\fontQ=cmr17
\font\fontR=cmssbx10
\font\bg=cmbx10 scaled 1200

\def\title#1{\centerline{\fontQ#1} \vskip 17mm}
\def\author#1{\centerline{\fontR#1} \vskip 6mm}
\def\address#1{\centerline{\it#1}\vskip 2mm}
\def\date#1{\centerline{#1}\vskip 60mm}
\def\abstract#1{{\centerline{\bf Abstract}} \vskip 3mm \par #1}

\def\references#1{{\leftline{\bg References}} \vskip 3mm
\baselineskip=13pt}
\def\preprintnumber#1{\rightline{\fontP #1}\vskip 23mm}
\def\ref[#1]#2{\item{[#1]}#2}

\def\chapter#1{\leftline{\bg#1} \vskip 5mm}
\def\section#1{\leftline{\bg#1} \vskip 5mm}
\def\appendix(#1,#2){{\leftskip=-5mm\noindent{\bg#1}\par}\noindent
{\bf #2} \vskip 3mm}
\def\endpage{\vfill \eject}

\font\fontD=cmr7

\def\footD[#1,#2]{\footnote{$^{\# #1}$}{\baselineskip=4mm
\fontD\hang\enskip\enskip #2\vskip -3mm\par}}

\def\varz{{\delta_\zeta}^{(n-{1\over 2})}}
\def\varhz{{{\hat \delta}_\zeta}^{(n-{1\over 2})}}
\def\halfm-#1{{#1-{1\over 2}}}
\def\halfp-#1{{#1+{1\over 2}}}
\def\mpair{(\Phi ,\, \Theta)}
\def\mdpair{(\Lambda ,\, \Theta)}
\def\mddpair{(\Lambda ,\, \Theta_{diag})}
\def\Thetad{\Theta_{diag}}
\def\Dvan{\Delta(\Lambda)}
\def\prodth{\prod_{i<j}\> \theta_{ij}\theta_{ji}}
\def\lam-#1#2{{\lambda_#1 -\lambda_#2}}
\def\lamn-#1{{\lambda_#1}^n}
\def\th-#1{\theta_{#1#1}}
\def\lmd-#1{\lambda_{#1}}
\def\sgm-#1{\sigma_{#1}}
\def\sym-#1{s_{#1}}
\footline={\hfill}
\preprintnumber{RIMS-867}
\title{Matrix Model with Superconformal Symmetry}
\author{Michiaki TAKAMA}
\address{Research Institute for Mathematical Sciences,}
\par\noindent
\address{Kyoto University, Kyoto, 606 Japan}
\date{February 1992}
\abstract{A matrix model is presented which leads to the discrete
``eigenvalue model'' proposed recently by Alvarez-Gaum\'e {\it et.al.}
for 2D supergravity (coupled to superconformal matters).}
\endpage
\footline={\hfill\ -- \folio\ -- \hfill}
\pageno=1
Through the recent advances in understanding 2D quantum gravity,
it has been confirmed that the nonperturbative formulation of the
theory is successfully defined in terms of matrix models[1].
In these models, the integral over matrices realizes the
random sum of the
triangulations of two dimensional surfaces. Encouraged by these
successes, we incline to proceed to the supersymmetric version of the
theory. However, in the case of discretized gravity,
its supersymmetrization perplexes us with the difficulty of imaging
what is meant by ``triangulations of super-surfaces''. In formulating
matrix models to get supersymmetry[2], most of the attempts that
have been made so far thus deal with the target space supersymmetry
and not the world-surface supersymmetry(, excepting phenomenological
approaches by means of the
super-extension of soliton equations[3]). \par
Very recently, Alvarez-Gaum\'e {\it et.al.}[4] proposed a discrete model
which seems probably to provide a discrete version of 2D supergravity.
Taking the ``planar'' limit, they showed that the model reproduces
the string susceptibility and the spectrum of anomalous
dimensions of the $(2\, ,\> 4m)$-minimal superconformal models
coupled to 2D supergravity.
Since their model is presented in
the form of ``eigenvalue model'', its interpretation such as
triangulated super-surfaces is impossible as it stands.
In this letter, we lift their model to a matrix model.
The resulting model, however, has a peculiar form which seems
not so easy to deal with. We have not yet had an answer
whether such an interpretation mentioned above is in fact possible.
\par
The idea to construct a model is very simple. As expounded
in ref.[4], the guiding principle is the super-Virasoro structure
of the partition function. Let us consider a super-pair of
$N \times N$
hermitian matrix $(\Phi ,\, \Theta)$, whose matrix elements
$\phi_{ij}$, $\theta_{ij}$\quad $1\leq i,j\leq N$ are Grassmann
even and odd variables respectively.\footD[1,The hermiticity
of $\scriptstyle{\Theta}$ would be imposed such that
$\scriptstyle{
\overline{\theta_{ij}\zeta}={\bar \zeta}\,\overline{\theta_{ij}}
=\theta_{ji}\zeta}$ with $\scriptstyle{\zeta}$ being a real
odd constant ($\scriptstyle{{\bar \zeta}= \zeta}$), hence
$\scriptstyle{\overline{\theta_{ij}}=-\theta_{ji}}$.
Throughout this letter, we deal with $\scriptstyle{\phi_{ij}}$ and
$\scriptstyle{\theta_{ij}}$ as independent $\scriptstyle{N^2+N^2}$
variables for simplicity.] When $\Phi$ is diagonal,
we denote it by
$\Phi = \Lambda =
\biggl({}^{{}^{\lambda_1}} \ddots_{{}_{\lambda_N}}\biggr)$.
In usual matrix models, Virasoro structure
can be recognized by the change of the variables
$1+{\delta_\varepsilon}^{(n)}\> : \> \Phi \longrightarrow
\Phi + \varepsilon \Phi^{n+1}$
and as a result the partition function obeys the Virasoro
constraints[5]. By analogy, we consider the transformation,
with an odd constant parameter $\zeta$:
$$
1+\varz \> :\> (\Phi ,\, \Theta)\> \longrightarrow
  (\, \Phi + \Theta \Phi^n \zeta ,\> \Theta +\Phi^n \zeta \, )
\quad .\eqno(1)
$$
The generators of the transformation is written as
${\tilde G}_{\halfm-n}\, =
-Tr\,\Phi^n\, (\partial_\Theta - \Theta \partial_\Phi)$ with
$(\partial_\Phi)_{ij}\> ,\>\, (\partial_\Theta)_{ij} =
\partial_{\Phi_{ji}}\> ,\>\, \partial_{\Theta_{ji}}$.
Obviously, only when $(\Phi ,\, \Theta)$ is guaranteed
in a proper way to satisfy the conditions
$[\,\Phi \, ,\, \Theta\, ]=0$
and $\Theta^2 =0$, the generators ${\tilde G}_{\halfm-n}\quad n\geq 0$
form the super-Virasoro algebra with
${\tilde L}_n\, =
-Tr(\,\Phi^{n+1}\,\partial_\Phi +
{{n+1}\over 2} \Theta \Phi^n \partial_\Theta\, )\quad n\geq -1$.
We define the partition function of our model as follows:
$$
Z_N=\int d\mu\mpair\, e^{-\beta\, Tr\, V\mpair},
\eqno(2)
$$
with
$$
d\mu\mpair =d^{N^2}\Phi\, d^{N^2}\Theta\> F_N\mpair\> \quad
d^{N^2}\Theta\equiv
\prod_{i=1}^{N}\> d\theta_{ii}\, \prod_{1\leq i<j \leq N}
d\theta_{ij}d\theta_{ji}\quad .
\eqno(3)
$$
{}From the above argument, we require the super-function $F_N\mpair$
to have the following properties:
 \item{(A)}\quad $F_N$ is invariant under the adjoint action
of the $U(N)$ group to $\mpair$, {\it i.e.}
$$
F_N\mpair = F_N({}^U\Phi\, , \, {}^U\Theta)\quad\>
({}^U\Phi\, , \, {}^U\Theta)=
(U\Phi U^{\dagger},\, U\Theta U^{\dagger})\quad .
$$
 \item{(B)}\quad Under the multiplication by $F_N$, it is ensured
that $[\,\Phi \, ,\, \Theta\, ]=0$
and $\Theta^2 =0$ in the integrand.
 \item{(C)}\quad $F_N$ is invariant under the transformation (1).
\par
{}From the requirement(A), the measure $d\mu\mpair$ becomes invariant
under the adjoint action of $U(N)$. As usual, the $U(N)$
integration yields the volume factor of the unitary group,
which we drop. The partition function (2) is then given by
the integral:
$$
\eqalign{
\int d^N\Lambda\, d^{N^2}\Theta\>  &\Delta(\Lambda)^2\,
F_N\mdpair\, e^{-\beta\, Tr\, V\mdpair},\cr
\Delta(\Lambda) &=\prod_{1\leq i<j \leq N}
(\lambda_i -\lambda_j)\cr}\quad .
\eqno(4)
$$
The requirement(B) and (C) are the sufficient conditions
for ${\tilde G}_\halfm-n$ and ${\tilde L}_{n-1}\quad (n\geq0)$
to form (the half of) the super-Virasoro algebra.
Although the potential $V$ is assumed to be a polynomial of the
matrices $\Phi$ and $\Theta$, only the terms up to the
linear order with respect to $\Theta$ contribute because of (B).\par
Let us start seeking the expression of $F_N$.
For (A) and (B), it is necessary and sufficient to find
a manifestly $U(N)$-invariant super-function $G_N\mpair$
such that
$$
G_N\mdpair \propto \prod_{i<j}\> \theta_{ij}\theta_{ji}\quad .
$$
We find the following as such a function:
$$
G_N\mpair = {1 \over {{N \choose 2}!}}
[\> Tr\, \Phi\Theta^2\> ]^{N \choose 2}\quad .
\eqno(5)
$$
In fact,
$$
G_N\mdpair = \Dvan \prodth\quad .
\eqno(6)
$$
Here we make some comments on $G_N$. The super-function $G_N$
is essentially $\delta^{N^2}(\Xi)=\prod_{i,j}\Xi_{ij}$, with
$\Xi =[\,\Phi \, ,\, \Theta\, ]$. This expression is manifestly
$U(N)$-invariant, but equals to zero in practice, since
there are $N$ linear relations among the $N^2$ variables
$\Xi_{ij}$. These linear dependency are described by
introducing $U(N)$-invariant odd variables
$\xi_1\, \cdots , \xi_N$ as follows:
$$
\xi_j=\Sigma_{j-{1 \over 2}}(\Phi\, ,\,\Xi)\equiv
{1 \over {j!}}\> (\partial_t)^j\partial_\zeta
\det\big\{\, I+t\, (\Phi + \zeta\,\Xi)\, \big\}\Big|_{t=0}\quad ,
\eqno(7)
$$
with $\zeta$ being an odd parameter.
In fact, if the commutator $[\,\Phi \, ,\, \Theta\, ]$ is
substituted for $\Xi$ in (7), these variables become
automatically zero. One finds
$$
\big(\prod_i\,\partial_{\xi_i}\big)\,
\prod_{i,j}\, \Xi_{ij}\>\,\Big|_{\Xi=[\,\Phi \, ,\, \Theta\, ]}
=G_N\mpair\quad.
\eqno(8)
$$
\par
Having regard to (4) and (6), we write $F_N\mdpair$ as
$$
F_N\mdpair = {1 \over \Dvan^3}\> f_N\mddpair\,\Dvan\,\prodth
\quad ,\eqno(9)
$$
where $\Thetad$ is a matrix of the diagonal part of $\Theta$,
{\it i.e.} $\Thetad =
\biggl({}^{{}^{\theta_{11}}} \ddots_{{}_{\theta_{NN}}}\biggr)$.
The requirement(A) can be established if we can rewrite
the factor ${1 \over \Dvan^3}\> f_N\mddpair$ to a
$U(N)$-invariant form. We postpone concerning this point,
and assume it for a while. As to the requirement(C), it is
sufficient to investigate the invariance at $\Phi =\Lambda$,
since
$$
F_N(\, \Phi + {\tilde \Theta} \Phi^n \zeta ,\> {\tilde \Theta}
+\Phi^n \zeta \, )
=F_N(\, \Lambda + \Theta \Lambda^n \zeta ,\> \Theta
+\Lambda^n \zeta \, )
$$
with $(\Lambda\, ,\,\Theta)=
({}^U\Phi\, , \, {}^U{\tilde \Theta})$,
provided that $F_N$ is $U(N)$-invariant.
For $\Lambda$ generic, {\it i.e.} $\lambda_i \not= \lambda_j$
for $i\not= j$, the matrix $\Lambda + \Theta \Lambda^n \zeta$ can be
diagonalized as follows
$$
(I+\Omega\,\zeta)^{-1}(\Lambda + \Theta \Lambda^n \zeta)
\>(I+\Omega\,\zeta)
=\Lambda + \Thetad \Lambda^n \zeta\quad ,
\eqno(10)
$$where
$$
{\Omega_{ij}=\left\{
\eqalign{
\hfil 0 \qquad &{\rm for}\quad i=j\cr
{{\theta_{ij}\lamn-j} \over \lam-ji}\quad
&{\rm for}\quad i\not= j\cr}
\right.}\quad .
\eqno(11)
$$
Hence we have
$$
F_N(\, \Lambda + \Theta \Lambda^n \zeta\> ,\>
\Theta +\Lambda^n \zeta \, )
=F_N\big(\, \Lambda + \Thetad \Lambda^n \zeta\> ,\>
\Theta +(\,\{\Omega\, ,\,\Theta\}+\Lambda^n\, )\zeta \, \big)
\eqno(12)
$$
and here
$$
\{\Omega\, ,\,\Theta\}_{ij}
=\sum_{k\, (\not= i)}
{{\theta_{ik}\theta_{kj}{\lamn-k}} \over {\lam-ki}}
-\sum_{k\, (\not= j)}
{{\theta_{ik}\theta_{kj}{\lamn-j}} \over {\lam-kj}}
\quad .
\eqno(13)
$$
Consequently, (C) is equivalent to the claim for $F_N\mdpair $
to be invariant under the following transformation:
$$
1+\varhz :\>\>\left\{
\eqalign{
\lambda_i &
\longrightarrow \lambda_i +\theta_{ii}\lamn-i\zeta\cr
\theta_{ij} &
\longrightarrow \theta_{ij}+
\{\Omega\, ,\,\Theta\}_{ij}+\delta_{ij}\lamn-i\zeta\cr}\right.
\quad .
\eqno(14)
$$
Then we see that
$$
\eqalignno{
\varhz &\prodth =
\big\{ \sum_{i<j}{(\lamn-i +\lamn-j)\over\lam-ij}\>
(\th-i -\th-j)\big\}\prodth\>\zeta\quad ,&(15)\cr
\varhz &\log\Dvan^{-2}=
-2\sum_{i<j}{(\lamn-i\th-i -\lamn-j\th-j)\over\lam-ij}\>\zeta
\quad .&(16)\cr}
$$
Accordingly, the invariance of $F_N$ requires
$$
\varhz\log f_N\mddpair=
\sum_{i<j}{(\lamn-i -\lamn-j)\over\lam-ij}\>
(\th-i +\th-j)\>\zeta\quad .
\eqno(17)
$$
Note that the function $f_N\mddpair$ is defined modulo $\theta_{ij}
\>\> (i\not= j)$. Taking this into account, the left hand side of
(17) is written as
$\zeta\,\sum_i \lamn-i (\partial_{\th-i}-\th-i\partial_{\lambda_i})
\log f_N$.
Hence we obtain
$$
-\sum_i \lamn-i (\partial_{\th-i}-\th-i\partial_{\lambda_i})
\log f_N\mddpair=
\sum_{i\not= j}{(\lamn-i -\lamn-j)\over\lam-ij}\>\th-i\quad .
\eqno(18)
$$
This is the same equation that led to the eigenvalue model
of ref.[4]. The equation (18) is easily integrated and
its unique solution (up to a multiplicative constant) is
$$
f_N\mddpair = \prod_{i<j}(\lmd-i -\lmd-j -\th-i\th-j)
\quad .\eqno(19)
$$
The condition(C) may seem rather stronger requirement.
In order to ensure the super-Virasoro structure, it suffices
that $\varhz F_N\mdpair$ becomes a product of $\prodth$ and a factor
which can be compensated by properly differentiating the integrand
$e^{-\beta\, Tr\,V}$ with respect to the coupling constants.
Note that, however, the expression in (17) takes the
form which can not be obtained by such differentiations for $n=0$. \par
We return to the requirement(A). From (19), we see that
$$
{1 \over \Dvan^3}\> f_N\mddpair =
{1 \over {\prod_{i<j}\big\{\> (\lmd-i -\lmd-j)^2 +
(\lmd-i -\lmd-j)\th-i\th-j\>\big\} }}\quad .
\eqno(20)
$$
The denominator of the right hand side
is a symmetric super-polynomial of the
variables $(\lmd-i\, , \,\th-i)\quad 1\leq i\leq N$, namely,
a super-polynomial invariant under the permutation
$(\lmd-i\, , \,\th-i) \leftrightarrow (\lmd-j\, , \,\th-j)$.
We can also define for the super case elementary symmetric
super-polynomials $\sgm-{{j \over 2}}\quad j=1,\cdots ,2N$:
$$
\eqalign{
{}&\sgm-r =\sgm-r (\Lambda)
\equiv {\rm usual\>\> elementary\>\> polynomial\>\> of}\>\>
r{\scriptstyle -}{\rm th\>\>
order}\quad ,\cr
{}&\sgm-{\halfm-r}\equiv
\sum_{i=1}^N\th-i\partial_{\lmd-i}\sgm-r
=\Sigma_{\halfm-r}\mddpair \>\> {\rm defined\>\> by}\>\> (7)
\quad ,\cr
{}&\qquad r=1,2,\cdots ,N\quad .\cr}
\eqno(21)
$$
Generalizing the argument of the usual case, one can
prove that any symmetric super-polynomial is expressed as a
super-polynomial in $\sgm-{{j\over 2}}$'s. Let us introduce
another type of symmetric super-polynomials defined by
$$
\sym-r =\sum_{i=1}^N \lmd-i^r\qquad {\rm and}\qquad
\sym-{\halfm-r}=\sum_{i=1}^N \th-i\lmd-i^{r-1}\quad .
\eqno(22)
$$
Then, one finds the generalized Newton's formulae:
$$
\eqalign{
{}&
n\sgm-n -\sgm-{n-1}\sym-1 + \cdots
+(-1)^r\sgm-{n-r}\sym-r +\cdots +(-1)^n\sym-n=0\cr
{}&
\sgm-{\halfm-n} -\sgm-{n-1}\sym-{{1\over 2}} + \cdots
+(-1)^r\sgm-{n-r}\sym-{\halfm-r} +\cdots +(-1)^n\sym-{\halfm-n}=0
\cr}\quad .
\eqno(23)
$$
These formulae enable us to write $\sgm-{{j\over 2}}$'s in terms
of $\sym-{{j\over 2}}$'s. We also remark
$\sym-r = Tr\, \Phi^r$ and $\sym-{\halfm-r}
=Tr\,\Thetad\Lambda^{r-1}=Tr\,\Theta\Lambda^{r-1}$.
{}From the above, we understand that the expression (20)
is rewritten as
$$
{1\over\Dvan^3}\> f_N\mddpair ={1\over {E_N\mpair}}
\Big|_{\Phi=\Lambda}\quad ,
\eqno(24)
$$
where $E_N\mpair$ is a certain super-polynomial in
$\sym-{{j\over 2}}\mpair$'s:
$$
E_N\mpair =\sum_{\sum_1^\mu j +\sum_1^{2\nu}k+\>\,\nu =N(N-1)}
c_{j_1 \cdots j_\mu ;k_1 \cdots k_{2\nu}}
\sym-{j_1}\cdots\sym-{j_\mu}\>
\sym-{\halfp-{k_1}}\cdots\sym-{\halfp-{k_{2\nu}}}
\quad .
\eqno(25)
$$
We have now reached the conclusion that the preceding requirements
(A), (B) and (C) determine uniquely the desired measure
$d\mu\mpair$ and that the partition function (2) is reduced to
that of the eigenvalue model in [4].
The explicit form of the model depends heavily
on the matrix size $N$ and seems somewhat ugly, especially due to
the factor $E_N\mpair^{-1}$. (In spite of the appearance,
the matrix integral is well defined, as can be seen from the
reduced integral over the eigenvalues.) For instance, in the
most simple case $N=2$, our model is the following:
$$
Z_2=\int d^4\Phi\>d^4\Theta\>\, {Tr\> \Phi\Theta^2 \over
{2\, Tr\,\Phi^2 -(Tr\, \Phi)^2+
Tr\, \Phi\Theta\cdot Tr\,\Theta}}\>e^{-\beta\,Tr\> V\mpair}\quad .
\eqno(26)
$$
The problem is whether the matrix integral of the model bears
the translation into the random sum of the triangulated
super-surfaces by means of the graphical expansion.
At this stage, it is far from obvious and needs
more study.
\par
We conclude this letter with a remark on the super-Virasoro
constraints. The partition function (2)
is obviously invariant under the shift of the variables (1).
Following ref.[4], let us write the potential $V$ as
$$
V\mpair =\sum_{k\geq 0}\> (\> g_k\Phi^k +
\xi_{\halfp-k}\Theta\Phi^k\> )\quad .
\eqno(27)
$$
The change of the integrand is then given by
$$
-\zeta\>\sum_{k\geq 0}\> (\> k\,g_k\,\partial_{\xi_{\halfm-n +k}}
+ \xi_{\halfp-k}\,\partial_{g_{n+k}}\> )\> e^{-\beta\,Tr\> V}
\quad .
\eqno(28)
$$
Because the function $F_N$ is invariant under the shift,
the change of the measure $d\mu$ comes only from the Jacobian:
$$
{\partial(\, \Phi + \Theta \Phi^n \zeta ,
\> \Theta +\Phi^n \zeta \, )\over\partial\mpair}
=1-\zeta\>\sum_{k=0}^{n-1}\> Tr\>\Theta\Phi^k\cdot Tr\>\Phi^{n-1-k}
\quad
\eqno(29)
$$
Hence the super-Virasoro constraints[4] on the partition function
follows:
$$
G_{\halfm-n}\>\> Z_N=0\qquad n\geq 0
\eqno(30)
$$
with
$$
G_{\halfm-n}=\sum_{k\geq 0}
(\> k\,g_k\,\partial_{\xi_{\halfm-n +k}}
+ \xi_{\halfp-k}\,\partial_{g_{n+k}} +{1\over \beta^2}\>
\sum_{k=0}^{n-1}\>\partial_{\xi_{\halfp-k}}\partial_{g_{n-1-k}}\>)
\quad .
\eqno(31)
$$
\vskip 3cm
\references
\noindent\hang\textindent{[1]}
E. Br\'ezin and V.A. Kazakov, {\it Phys. Lett.}
{\bf 236B}(1990)144;
\item{}M.R. Douglas and S.H. Shenker, {\it Nucl. Phys.}
{\bf B335}(1990)635;
\item{}D.J. Gross and A.A. Migdal, {\it Nucl. Phys.}
{\bf B340}(1990)333.
\ref[2]J. Alfaro and P.H. Damgaard, {\it Phys. Lett.}
{\bf 222B}(1989)429;
\item{}A. Mikovi\'c and W. Siegel, {\it Phys. Lett.}
{\bf 240B}(1990)363;
\item{}S. Bellucci, T.R. Govindrajan, A. Kumar and R.N. Oerter,
\hfill\break {\it Phys. Lett.} {\bf 249B}(1990)49;
\item{}J. Ambj\o rn and S. Varsted, {\it Phys. Lett.}
{\bf 257B}(1991)305;
\item{}E. Marinari and G. Parisi, {\it Phys. Lett.}
{\bf 240B}(1990)375;
\item{}M. Karliner and A.A. Migdal, {\it Mod. Phys. Lett.}
{\bf A5}(1990)2565;
\item{}A. Dabholkar, Rutgers preprint RU-91-20(1991);
\item{}J. Gonzalez, {\it Phys. Lett.} {\bf 255B}(1991)367;
\item{}S. Nojiri, {\it Prog. Theor. Phys.} {\bf 85}(1991)671;
\item{}A. Jevicki and J.P. Rodrigues, {\it Phys. Lett.}
{\bf 268B}(1991)53.
\item{}\qquad Supermatrix models were studied in
\item{}G. Gilbert and M. Perry, {\it Nucl. Phys.} {\bf B364}(1991)734;
\item{}L. Alvarez-Gaum\'e and J.L. Ma\~nes, {\it Mod. Phys. Lett.}
{\bf A6}(1991)2039;
\item{}S.A. Yost, U. Florida preprint UFIFT-HEP-91-12.
\ref[3]P. Di Francesco, J. Distler and D. Kutasov,
{\it Mod. Phys. Lett.} {\bf A5}(1990)2135;
\item{}M. Awada, U. Florida preprints UFIFT-HEP-90-18 and -29.
\ref[4]L. Alvarez-Gaum\'e, H. Itoyama, J.L. Ma\~nes and
A. Zadra, ``Superloop Equations and Two Dimensional
Supergravity'', CERN preprint CERN-TH.6329/91.
\ref[5]A. Mironov and A. Morozov, {\it Phys. Lett.}
{\bf 252B}(1990)47;
\item{}Y. Matsuo, unpublished.
\item{}\qquad As to the Virasoro constraints after taken the
double scaling limit, see
\item{}M. Fukuma, H. Kawai and R. Nakayama,
{\it Int. J. Mod. Phys.} {\bf A6}(1991)1385;
\item{}R. Dijkgraaf, E. Verlinde and H. Verlinde,
{\it Nucl. Phys.} {\bf B348}(1991)435.

\end